\documentclass[12pt]{article} 
\usepackage[xdvi]{graphicx} 
\usepackage{latexsym}
\usepackage{amssymb}
\usepackage{amsmath} 
\usepackage{subfigure}
\usepackage{epsfig}
\usepackage{array}
\usepackage{cite} 
\usepackage{psfrag}

\begin{document}   
\newcommand{\todo}[1]{{\em \small {#1}}\marginpar{$\Longleftarrow$}}   
\newcommand{\labell}[1]{\label{#1}\qquad_{#1}} 


\begin{flushright}
DCPT--03/53
\end{flushright}

\bigskip
\bigskip
\begin{center}
   {\Large \bf Using D-Strings to Describe }

\bigskip

{\Large \bf Monopole Scattering }
   \end{center}

\bigskip
\bigskip

\centerline{\bf Jessica K. Barrett$^{a,b,c\,}$\footnote{Jessica.Barrett@esi.ac.at}, Peter Bowcock$^{a}$\footnote{peter.bowcock@durham.ac.uk}}

\bigskip

\centerline{$^a$ \it Centre for Particle Theory} \centerline{\it
  Department of Mathematical Sciences} \centerline{\it University of
  Durham} \centerline{\it Durham, DH1 3LE, U.K.}

\bigskip

\centerline{$^b$ \it Erwin Schr\"odinger Institute} \centerline{\it Boltzmanngasse 9} \centerline{\it A-1090 Vienna} \centerline{\it Austria}

\bigskip

\centerline{$^c$ \it Institut für Theoretische Physik} \centerline{\it Technische Universität Wien} \centerline{\it Wiedner Hauptstr. 8-10/136} \centerline{\it A-1040 Vienna} \centerline{\it Austria}


\bigskip
\bigskip


\begin{abstract}
We describe the scattering of D-strings stretched between D3-branes, working from the D-string perspective. From the D3-brane perspective the ends of the D-strings are magnetic monopoles, and so the scattering we describe is equivalent to monopole scattering. Our aim is to test the prediction made by Manton for the energy radiated during monopole scattering.
\end{abstract}

\newpage \baselineskip=18pt \setcounter{footnote}{0}

\section{Introduction}

Since the discovery of D-branes in string theory, and the realisation that they can be alternatively interpreted as soliton solutions of supergravity, it has been apparent that our knowledge of D-branes is crucial for our understanding of string theory and M-theory. See ref. \cite{Johnson:2003gi} for an review of the known properties of D-branes.

An interesting D-brane configuration to study is a bunch of F- and/or D-strings attached to a D3-brane. The F-string case, known as the BIon spike solution, was studied extensively in refs. \cite{Callan:1998kz}, \cite{Gibbons:1998xz}, \cite{Howe:1998ue}. The D-string case has also received much attention in the literature \cite{Diaconescu:1997rk}, \cite{Constable:1999ac}, \cite{Constable:2001ag}. This configuration is particularly interesting, because it can be studied either from the D3-brane perspective, or from the D-string perspective, by using the Born-Infeld action with the appropriate dimension in each case. From the D3-brane perspective the end of the $k$ D-strings is a charge $k$ magnetic monopole in the (3+1)-dimensional D3-brane worldvolume. It is described by the usual Bogomol'nyi equations for a magnetic monopole, i.e.
\begin{equation}
B_i = D_i \Phi \,\, ,
\end{equation}
where $i = 1,\ldots,3$ are the D3-brane's spatial directions. Here $B_i$ is the magnetic field in the D3-brane's worldvolume, $\Phi$ is an excited transverse field, and $D_i$ is the covariant derivative taken with respect to the gauge field living on the D3-brane worldvolume. The configuration has also been studied from the D-string perspective in refs. \cite{Constable:1999ac} and \cite{Constable:2001ag}. The end of the D-strings is still a monopole (as we would expect), but this time it is described by Nahm's equations,
\begin{equation}
\partial_{\sigma} \Phi^i + \frac{1}{2i} \epsilon_{ijk} [\Phi^j,\Phi^k] = 0 \,\, ,
\label{eq:Nahms-equations}
\end{equation} 
where $\sigma$ is the D-strings' spatial direction, and the $\Phi^i$ are three excited transverse fields on the D-strings' worldvolume (here we have taken the $\Phi^i$ to be Hermitian matrices). So the D-string perspective provides a physical realisation of the (1+1)-dimensional description of a monopole, the ADHMN construction. See ref. \cite{Corrigan:1984sv} for a review of the ADHMN construction. See also refs. \cite{Tsimpis:1998zh}, \cite{Chen:2002vb} for discussions of how the boundary conditions on the Nahm data arise in the D-string picture.
In this paper we will study this configuration, with D-strings stretched between parallel D3-branes, working from the D-string perspective. 


Much research has been done concerning the properties of monopoles in $(3+1)$ dimensions - see, for example, refs. \cite{Atiyah:1988jp}, \cite{Sutcliffe:1997ec} and \cite{Harvey:1996ur} for reviews. So it seems natural to ask whether there is any monopole calculation, whose result is known, which could be recalculated in the $(1+1)$ dimensional D-string picture. We could then compare the two results. One possibility is the calculation of the energy radiated during scattering of the D-strings, which has been calculated for the BPS magnetic monopole case by Manton and Samols in ref. \cite{Manton:1988bn}. They obtained the result $E_{\rm{rad}} \sim 1.35 m_{\rm{mon}} v_{\rm{init}}^5$, where $m_{\rm{mon}}$ is the mass of the monopoles, and $v_{\rm{init}}$ is their initial velocity. We will aim to recalculate this result, this time working from the point of view of the D-strings. In this paper we present our analytic calculations, which lead to equations of motion for the non-zero mode perturbations of the D-strings. We will present our result for the energy radiated, which we have calculated numerically, in a later paper.

In order to work from the D-string perspective we will need to use the action for two parallel D-strings, which is the non-abelian Born-Infeld action. Up to order $\alpha'^2$ and order $F^6$ this action is given by the usual abelian Born-Infeld action with an additional symmetrised trace acting on all the fields. This symmetrised trace description was first introduced by Tseytlin in ref. \cite{Tseytlin:1997cs}, and was discussed further in refs. \cite{Tseytlin:1999dj} and \cite{Myers:1999ps}. This description will be sufficient for our purposes, since we will be expanding the action as a series in $\alpha'$ and we will set the gauge field to zero. For discussions about higher order corrections to the non-abelian Born-Infeld action see, for example, refs. \cite{Bilal:2001hb}, \cite{Koerber:2002zb} and \cite{Bergshoeff:2000cx}.

In order to calculate the energy radiated during the scattering of the D-strings we will need to analyse the perturbations of the D-strings. There has been some work done on this subject in the literature. For example, \cite{Callan:1998kz}, \cite{Lee:1998xh} and \cite{Kastor:1999ag} calculate the perturbations of the BIon spike, working from the D3-brane perspective. The perturbations of the D-string have been studied in \cite{Savvidy:1999wx}, again working from the D3-brane perspective. In \cite{Constable:1999ac} the perturbations have been studied from the D-string point of view for the case of a semi-infinite spike attached to a D3-brane. In ref. \cite{Hamanaka:2001pe} the scattering of D0-branes on D4-branes was discussed, which is T-dual to the D1/D3-brane case considered here. See also ref. \cite{Rey:1997sp} for a discussion of the dynamics of the triple string junction, and ref. \cite{Rey:1998ik} for a discussion of BIon spike perturbations from the point of view of the AdS/CFT correspondence.  
   
As a caveat, recall that the Born-Infeld action is an approximate action, which
becomes unreliable when spacetime is highly curved. Since the geometry as the
D-strings funnel out becomes highly curved, it is possible that this may affect
our results. Nevertheless, the calculation remains interesting.

The layout of the paper is as follows. In section \ref{sec:energy_rad} we review the calculation of the energy radiated during monopole scattering given in ref.\cite{Manton:1988bn}. In section \ref{sec:BI_flat} we analyse the non-abelian Born-Infeld action for D-strings in flat background, and present the solution corresponding to D-strings stretched between D3-branes. We point out that it is necessary to be careful when taking the $\alpha^{\prime} \to 0$ limit to ensure that the mass of the W-boson, and the mass of the monopole remain finite. We also discuss the validity of the solutions, given the limitations of the Born-Infeld action described above. In section \ref{sec:BI_D3brane} we repeat the calculation from section \ref{sec:BI_flat}, but this time with a D3-brane background. In section \ref{sec:scattering} we describe the scattering of the two D-strings using the slow motion technique of Manton (ref. \cite{Manton:1982mp}). In section \ref{sec:perturbation} we examine the effects of perturbing the BPS solution, and calculate the  equations of motion for the perturbations. We present our conclusions in section \ref{sec:conclusions}.

\section{The Energy Radiated During Monopole Scattering}
\label{sec:energy_rad}

In this section we briefly review the work of Manton and Samols in ref. \cite{Manton:1988bn}, where they calculate the energy radiated during the scattering of BPS magnetic monopoles in the slow velocity limit. We take the case here in which the monopoles motion is restricted to the $x^1\!-\!x^2$ plane. They approach each other along the $x^1\!$-axis, then scatter at $90^{\circ}$ to move away from each other along the $x^2\!$-axis.

Since the monopoles are moving in the slow velocity limit, Manton and Samols were able to perform a multipole expansion of the monopole fields. The leading order multipole is the charge two monopole, which is invariant, and therefore does not contribute to the radiation. Since the monopoles are purely magnetic, there is no dipole contribution. So the leading order contribution to the energy radiated comes from the time-dependent quadrupole moment, which by symmetry can be written as:
\begin{equation}
\label{eq:quad_asymp}
Q_{ij} = Q_i \delta_{ij} \quad \textrm{(no sum over $i$)}  \,\, , \quad Q_1 + Q_2 + Q_3 = 0 \,\, .
\end{equation} 
Higher order multipoles contribute terms of higher orders in the velocity, and so can be neglected.

In the asymptotic limit, when the monopoles are far apart, they can be treated as pointlike objects. Then
\begin{displaymath}
Q_1 = -m_{\rm{mon}} r^2 \,\, , \quad Q_2 = \frac{1}{2}m_{\rm{mon}} r^2 \,\, , \quad Q_3 = \frac{1}{2} m_{\rm{mon}} r^2 \,\, ,
\end{displaymath}
for $t<0$, and with $Q_1$ and $Q_2$ interchanged for $t>0$. Here $m_{\rm{mon}}$ is the mass of each monopole. Also $r$ is the distance between the monopoles, which can be calculated in the asymptotic limit by integrating the Taub-NUT metric with respect to time to get
\begin{displaymath}
r(t) = 2vt + \ln{2vt} + c + \frac{\ln{2vt}}{2vt} + \frac{1}{4vt} \left(2c-1\right) + \cdots \,\, ,
\end{displaymath}
where $c$ is a constant with
\begin{displaymath}
c=1+\ln{2} \,\,.
\end{displaymath}
A constant $R$ is defined such that \eqref{eq:quad_asymp} is a good approximation to the quadrupole moments, providing $r>R$. A comparison with the Atiyah-Hitchin metric shows that it is safe to take $R \simeq 10$.

When the monopoles coincide the configuration is axisymmetric in the $x^1\!-\!x^2$ plane, and symmetry considerations imply,
\begin{equation}
\label{eq:quad_coin}
Q_1^0 = Q_2^0 = -\frac{1}{2} m_{\rm{mon}} \pi^2 \,\, , \quad Q_3^0 = m_{\rm{mon}}\pi^2 \,\, .
\end{equation}  
(see ref. \cite{Manton:1988bn} for the details of this calculation).

The behaviour of the quadrupole moment during the scattering is some smooth interpolation between \eqref{eq:quad_asymp} and \eqref{eq:quad_coin}. The total energy radiated, expressed in terms of $Q$ is given by
\begin{displaymath}
E_{\rm{rad}} = \frac{1}{432\pi} \sum_{i} \int \! dt \,\, \dddot{Q}^2_i(t) \,\,.
\end{displaymath}
Minimising this energy gives a lower bound for the energy radiated, which is
\begin{equation}
\label{eq:energy_rad}
E_{\rm{rad}} \simeq 1.35\, m_{\rm{mon}} \, v_{\rm{init}}^5 \,\, ,
\end{equation}
where $v_{\rm{init}}$ is the initial asymptotic velocity of each monopole. Moreover, Manton and Samols argue that \eqref{eq:energy_rad} is a good approximation to the energy radiated, not just a lower bound, because the minimising configuration they found exhibited all the important properties expected of the true evolution in time.

\section{The Action for D-Strings in Flat Background}
\label{sec:BI_flat}

In this section we investigate the non-abelian Born-Infeld action for D-strings in flat background, and obtain the solution corresponding to two D-strings stretched between two D3-branes.

\subsection{The Born-Infeld Action}

The non-abelian Born-Infeld action for D-strings is \cite{Myers:1999ps}
\begin{equation}
\label{eq:BIaction1}
S = -T_{1} \iint \! d\tau d\sigma \,\, \textrm{STr} \left( \frac
{e^{-\phi} \left( -D \right)^{1/2}}{\left( - \det
\left(E^{ij}\right)\right)^{1/2}} \right) \,\, ,
\end{equation}
where
\[
D = \det \begin{pmatrix}
E_{ab}-E_{ai}E^{ij}E_{jb}+\alpha^{\prime} F_{ab} & E_{ak}E^{kj}+\alpha^{\prime} D_{a}
\Phi^{j}\\
-E^{ik}E_{kb}-\alpha^{\prime} D_{b} \Phi^{i} & E^{ij}+\alpha^{\prime} \left[ \Phi^{i}
, \Phi^{j} \right]
\end{pmatrix} \,\, ,
\] 
with $E_{\mu\nu} = G_{\mu\nu}+B_{\mu\nu}$ and where $a,b=0,9$ are the D-string
directions which we have taken as $x^{0} = \tau = t$ and $x^{9} = \sigma$. Here 
$i,j=1,\ldots,8$ are the transverse directions, and we have taken $\Phi^{i} =
X^{i}/\alpha'$ to be the transverse fields. We also take the dilaton
$\phi$ to be constant, $B_{ij}=0$, and we choose the gauge such that $F_{ab}=0$.
We excite the fields $\Phi^{1}$, $\Phi^{2}$, $\Phi^{3}$, which will correspond
to the D3-brane directions, and we set $\Phi^{4} = \cdots = \Phi^{8} = 0$, which
is consistent with the equations of motion.

In flat background we have
\[
E_{\mu\nu} = \textrm{diag} \, (-1,1,1,1,1) \,\, ,
\]
giving
\[
\det(E^{ij})=-1 \,\, ,
\]
So
\begin{equation}
\label{eq:det1}
D = -\det \left( \begin{array}{ccccc}
-1 & 0 & \alpha' \dot{\Phi}_{1} & \alpha' \dot{\Phi}_{2} 
& \alpha' \dot{\Phi}_{3} \\
0 & 1 & \alpha' \Phi'_{1} & \alpha' \Phi'_{2} & \alpha' \Phi'_{3} \\
-\alpha' \dot{\Phi}_{1} & -\alpha' \Phi'_{1} & 1 & \alpha' [\Phi_{1},\Phi_{2}]
& \alpha' [\Phi_{1},\Phi_{3}] \\
-\alpha' \dot{\Phi}_{2} & -\alpha' \Phi'_{2} & \alpha' [\Phi_{2},\Phi_{1}] & 1
& \alpha' [\Phi_{2},\Phi_{3}] \\
-\alpha' \dot{\Phi}_{3} & -\alpha' \Phi'_{3} & \alpha' [\Phi_{3},\Phi_{1}]
& \alpha' [\Phi_{3},\Phi_{2}] & 1 
\end{array} \right) \,\, ,
\end{equation}
where a dot denotes differentiation with respect to $t$ and a prime denotes
differentiation with respect to $\sigma$.

To describe the D-strings funnelling out into D3-branes we should take the
transverse fields to belong to representations of the group $SU(2)$. In order to have two D-strings the appropriate representation to take is
the $(2\times2)$ representation, i.e. the Pauli matrices $\sigma_{j}$. So we take the following ansatz, which will correspond to the $90^{\circ}$ scattering of the D-strings,  
\begin{equation}
\label{eq:Phi}
\Phi_{j} = - \frac{1}{2} \phi_{j} \sigma_{j} \quad \textrm{(no summation over j)} \,\, ,
\end{equation}
where $j =1,2,3$, and the $\phi_{j}$ are real functions of $t$ and $\sigma$ (this ansatz is 
consistent with the $\Phi_{j}$ being Hermitian).

Substituting the ansatz \eqref{eq:Phi} into
the determinant \eqref{eq:det1}, then evaluating the determinant and performing the symmetrised trace gives the action
\begin{eqnarray}
S \!\! & = -T_{1} \int_{-\infty}^{\infty} \! dt \int_{0}^{L}  d\sigma & \!\!\! \bigg\{  
\Big(1 + \frac{\alpha'^{2}}{4}\partial_{\sigma} (\phi_{1}\phi_{2}\phi_{3})
\Big)^{2}
 + \frac{\alpha'^{2}}{4} \Big( (\phi'_{1} - \phi_{2}\phi_{3})^{2} + \cdots \Big) \nonumber \\
&& - \frac{\alpha'^{4}}{16} \Big( \partial_{t} (\phi_{1}\phi_{2}\phi_{3}) 
 \Big)^{2} 
 - \frac{\alpha'^{2}}{4} \Big( \dot{\phi}_{1}^{2} + \dot{\phi}_{2}^{2} 
 + \dot{\phi}_{3}^{2} \Big)
\nonumber \\ \label{eq:BIaction2}
&& - \frac{\alpha'^{4}}{16} \Big( (\dot{\phi}_{1} \phi'_{2} -  \dot{\phi}_{2} 
 \phi'_{1})^{2} + \cdots \Big) 
 \bigg\} ^{1/2} \,\, ,
\end{eqnarray}
where $L$ is the length of the string. In \eqref{eq:BIaction2} $+ \cdots$ denotes addition of cyclic permutations of $\phi_1$, $\phi_2$ and $\phi_3$ - we will adopt this notation throughout the rest of this paper.

\subsection{Taking the Low Energy Limit}
\label{sec:low_energy}

We will investigate what happens to the action \eqref{eq:BIaction2} in the low energy limit $\alpha' \to 0$. 

To ensure that the limit is taken in an appropriate manner we must make precise the dictionary between the monopole of Yang-Mills theory, and the monopole described by a D-string stretched between two D3-branes. The monopole of Yang-Mills theory is described by the usual Yang-Mills action in $(3+1)$ dimensions:
\begin{displaymath}
S_{\rm{YM}} = \frac{1}{g_{\rm{YM}}^2} \, \int \! dt\, d^3\!x \left( -\frac{1}{4}F_{\mu\nu}F^{\mu\nu} + \frac{1}{2} D_{\mu}\Phi D^{\mu}\Phi \right) \,\, ,
\end{displaymath}
where
\begin{displaymath}
D_{\mu} \Phi = \partial_{\mu} \Phi + i [A_{\mu},\Phi] \,\, ,
\end{displaymath}
and $g_{\rm{YM}}$ is the Yang-Mills coupling. Giving the Higgs field $\Phi$ an expectation value $v$ results in a mass for the W-boson with
\begin{equation}
\label{eq:W_mass}
m_{\rm{W}} = v \,\, .
\end{equation}
The monopole solution for the resulting action has mass
\begin{equation}
\label{eq:mon_mass}
m_{\rm{mon}} = \frac{v}{g_{\rm{YM}}^2} \,\, .
\end{equation}
We wish to compare the results \eqref{eq:W_mass} and \eqref{eq:mon_mass} to those of the correspnding D-brane picture, in which the monopole is represented by a D-string stretched between two D3-branes. The $(3+1)$-dimensional picture, equivalent to the monopole picture described above, is described by the $(3+1)$-dimensional Born-Infeld action for the D3-branes. In the D3-brane picture we have $g_{\rm{YM}}^2 = g_S$, where $g_S$ is the string coupling. The Higgs field $\Phi$ is represented by an excited transverse field, $X^9$ say, which represents the position of the D3-brane in the corresponding direction $x^9$. By dimensional analysis we have
\begin{displaymath}
v = \langle \Phi \rangle = \frac{\langle X^9 \rangle}{\alpha'} = \frac{L}{\alpha'} \,\,\, ,
\end{displaymath}
where $L$ represents the distance between the D3-branes in the $x^9$ direction, and therefore the length of the strings stretched between the D3-branes in that direction. The W-boson corresponds to a fundamental string stretched between the D3-branes, so we have
\begin{displaymath}
m_{\rm{W}} = T_{\rm{F}} L = \frac{L}{\alpha'} = v \,\,\, ,
\end{displaymath}
where $T_{\rm{F}}$ is the fundamental string tension, which agrees with the monopole result \eqref{eq:W_mass}. Also the monopole corresponds to a D-string stretched between the D3-branes, so
\begin{displaymath}
m_{\rm{mon}} = T_1 L = \frac{L}{g_S \alpha'} = \frac{v}{g_{\rm{YM}}^2} \,\,\, ,
\end{displaymath}
which also agrees with the monopole result \eqref{eq:mon_mass}.

Returning to the D-string picture, in taking the limit $\alpha' \to 0$ the usual procedure is to take the string length $L=\alpha' v$ to be fixed, while the Higgs expectation value $v \to \infty$. Applying this limit to the action \eqref{eq:BIaction2}, and keeping terms of order $\alpha'^2$ results in the Yang-Mills action expressed in terms of Nahm fields, as we would expect from ref. \cite{Witten:1996im},
\begin{eqnarray}
S & \!\! = -T_1 \, \int_{-\infty}^{\infty} \! dt \, \int_{0}^{L} d\sigma \,\, \bigg(\!\!\!\! & 1 - \frac{\alpha'^2}{4} \partial_{\sigma} \left(\phi_1\phi_2\phi_3\right) - \frac{\alpha'^2}{8} \left( (\phi'_1 - \phi_2\phi_3)^2 + \cdots \right) 
\nonumber \\ \label{eq:YMaction}
&& + \frac{\alpha'^2}{8} \left(\dot{\phi}_1^2 + \dot{\phi}_2^2 + \dot{\phi}_3^2 \right) \bigg) \,\, .
\end{eqnarray}  
The Bogomol'nyi equations for the action \eqref{eq:YMaction} are Nahm's equations \eqref{eq:Nahms-equations} with the ansatz \eqref{eq:Phi}. They are
\begin{equation}
\label{eq:nahm}
\phi'_1  =  \phi_2\phi_3  \,\, , \qquad
\phi'_2  =  \phi_3\phi_1  \,\, , \qquad
\phi'_3  =  \phi_1\phi_2  \,\, .
\end{equation}
In general, the equations of motion derived from \eqref{eq:YMaction} are
\begin{eqnarray}
\label{eq:phi1eqn}
\ddot{\phi}_1 - \phi''_1 + \phi_1(\phi_2^2 + \phi_3^2) & = & 0  \,\, ,\\
\label{eq:phi2eqn}
\ddot{\phi}_2 - \phi''_2 + \phi_2(\phi_3^2 + \phi_1^2) & = & 0  \,\, ,\\
\label{eq:phi3eqn}
\ddot{\phi}_3 - \phi''_3 + \phi_3(\phi_1^2 + \phi_2^2) & = & 0  \,\, .
\end{eqnarray}
However in this limit, i.e. as $v \to \infty$, $m_{\rm{W}}$ \eqref{eq:W_mass} and $m_{\rm{mon}}$ \eqref{eq:mon_mass} both blow up to infinity, which is clearly undesirable. 

Instead of taking the limit $\alpha' \to 0$ as described above, we will take an alternative limit in which we keep $v$ fixed and finite, while allowing the string length $L \to 0$. This limit ensures that the mass of the W-boson \eqref{eq:W_mass}, and the mass of the monopole \eqref{eq:mon_mass} remain fixed, unlike the previous limit we described. In the remainder of this section we will investigate the form of the action \eqref{eq:BIaction2} under this alternative limit.

\subsection{Rescaling the String Coordinate}
\label{sec:rescaling}

Although we are taking the limit in which $L\to0$, it will be easier in what follows to work in coordinates in which the
string length runs from 0 to 2. And so we perform the rescaling
\begin{equation}
\label{eq:rescale_sigma}
\sigma  \! \to  \! \xi \, \equiv   \frac{2}{\alpha'v} \sigma \,\, ,  
\end{equation}
To ensure that the Bogomol'nyi equations of the rescaled action retain
a familiar form, we must also rescale the $\phi_i$ as follows
\begin{equation}
\label{eq:rescale_phi}
\phi_{i}  \! \to \! f_{i}   \equiv  \frac{\alpha'v}{2} \phi_{i} \,\, .
\end{equation} 
Under this rescaling the
action \eqref{eq:BIaction2} becomes 
\begin{eqnarray}
\label{eq:BIaction3}
S \!\! & = -T_{1} \, \frac{\alpha'v}{2} \, \int_{-\infty}^{\infty} \! dt \, \int_{0}^{2} d\xi \,\,
\bigg\{ \!\!\!\! & \left( 1 + \frac{4}{\alpha'^{2}v^{4}} \partial_{\xi} (f_{1}f_{2}f_{3})
\right) ^{2}
+\frac{4}{\alpha'^{2}v^{4}} \left( (f'_{1} - f_{2}f_{3})^{2} + \cdots
\right) \nonumber\\
&& \qquad
-\frac{4}{\alpha'^{2}v^{6}} \left( \partial_{t} (f_{1}f_{2}f_{3}) 
\right) ^{2}
 - \frac{1}{v^{2}}\left( \dot{f}_{1}^{2} + \dot{f}_{2}^{2} 
+ \dot{f}_{3}^{2} \right) \nonumber \\
&& - \frac{4}{\alpha'^{2}v^{6}} \left( (\dot{f}_{1} f'_{2} - \dot{f}_{2} 
f'_{1} ) ^{2} + \cdots \right) \bigg\} ^{1/2} \,\,\, .
\end{eqnarray}

\subsection{The Bogomol'nyi Equations and the D3-Brane Solution}
\label{sec:bogomolnyi}

We define
\begin{equation}
\label{eq:H_tilde}
\tilde{H} = 1 + \frac{4}{\alpha'^{2}v^{4}} \partial_{\xi} 
(f_{1}f_{2}f_{3}) \,\,.
\end{equation}
Then the action \eqref{eq:BIaction3} becomes
\begin{align}
S = -T_{1} \frac{\alpha'v}{2} \int_{-\infty}^{\infty} \!\! dt \int_{0}^{2} \! d\xi & \,\,
\tilde{H} \left\{ 1 + \frac{4}{\alpha'^{2}v^{4}\tilde{H}^{2}} \left( 
(f'_{1} - f_{2}f_{3})^{2} + \cdots \right) \right. \nonumber \\
 & -\frac{4}{\alpha'^{2}v^{6}\tilde{H}^{2}} \left( \partial_{t} 
(f_{1}f_{2}f_{3}) \right) ^{2} 
 - \frac{1}{v^{2}\tilde{H}^{2}}\left( \dot{f}_{1}^{2} + \dot{f}_{2}^{2} 
+ \dot{f}_{3}^{2} \right) \nonumber \\
& \left. - \frac{4}{\alpha'^{2}v^{6}\tilde{H}^{2}} \left( (\dot{f}_{1} f'_{2} - 
\dot{f}_{2} f'_{1} ) ^{2} + \cdots \right) \right\} ^{1/2} \,\, .
\label{eq:BIaction4} 
\end{align}

For a static solution $\dot{f}_{1} = \dot{f}_{2} = \dot{f}_{3} = 0$, and
the action is minimised providing that
\begin{equation}
\label{eq:fprime}
f'_{1} = f_{2}f_{3} \,\, , \qquad
f'_{2} = f_{3}f_{1} \,\, , \qquad
f'_{3} = f_{1}f_{2} \,\, .
\end{equation}
When \eqref{eq:fprime} holds, the Lagrangian density is 
equal to a total derivative term. So \eqref{eq:fprime} are
the Bogomol'nyi equations for the action \eqref{eq:BIaction4}, and are identical to the Bogomol'nyi equations found by taking the usual limit \eqref{eq:nahm}, as was predicted in \cite{Hashimoto:1998px}. So for the purposes of finding the BPS solutions it does not matter which limit we take. 
However, since we will be interested here in perturbations of the BPS solutions, it is important that we use the correct limit for our calculations, i.e. we must use \eqref{eq:BIaction4} instead of \eqref{eq:YMaction}.

\begin{figure}
\centering
\psfrag{xi}{$\xi$}
\psfrag{f1}{$f_1$}
\psfrag{f2}{$f_2$}
\psfrag{f3}{$f_3$}
\mbox{
      \subfigure[]{\epsfig{figure=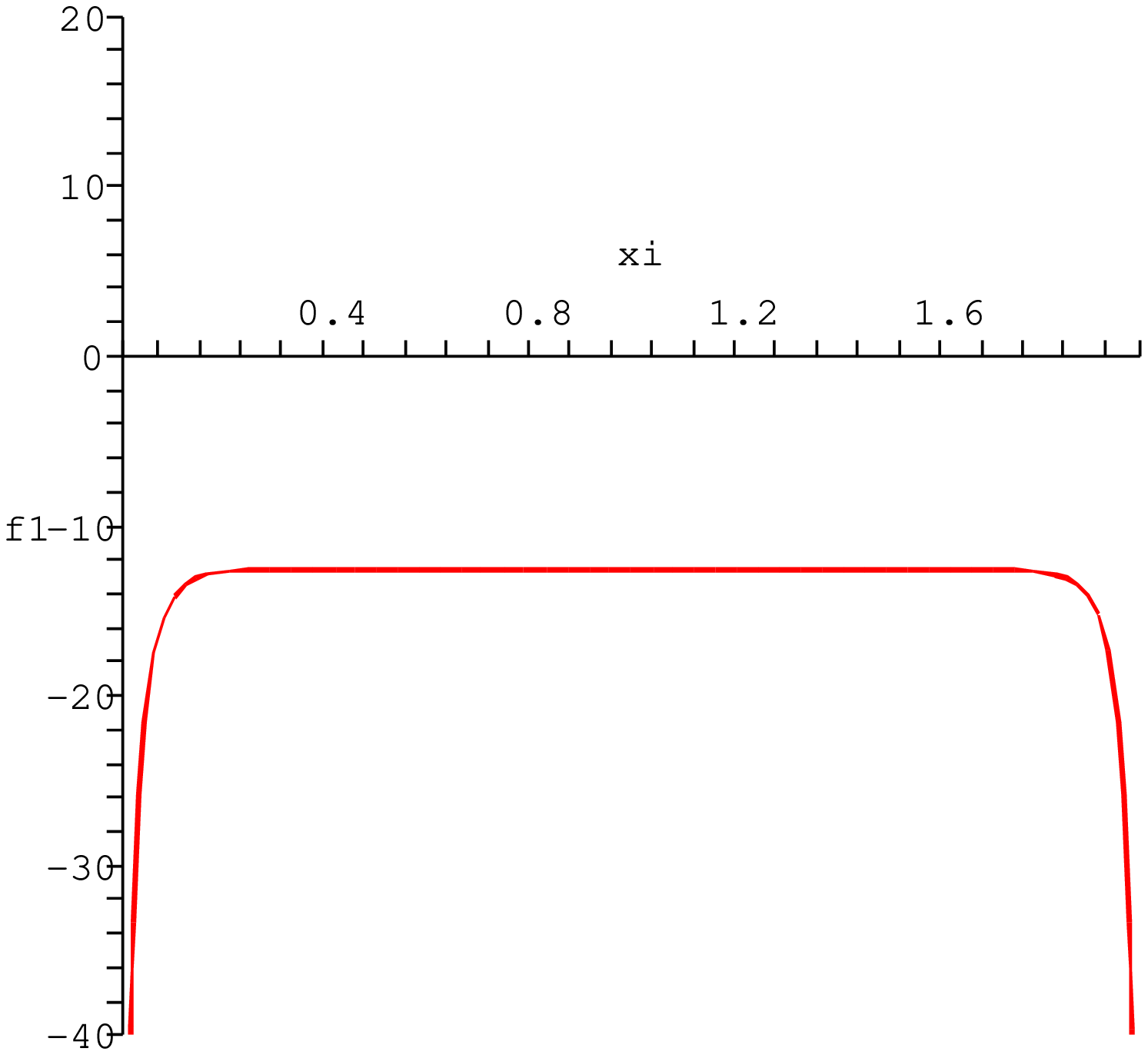,width=.35\textwidth}}
      \subfigure[]{\epsfig{figure=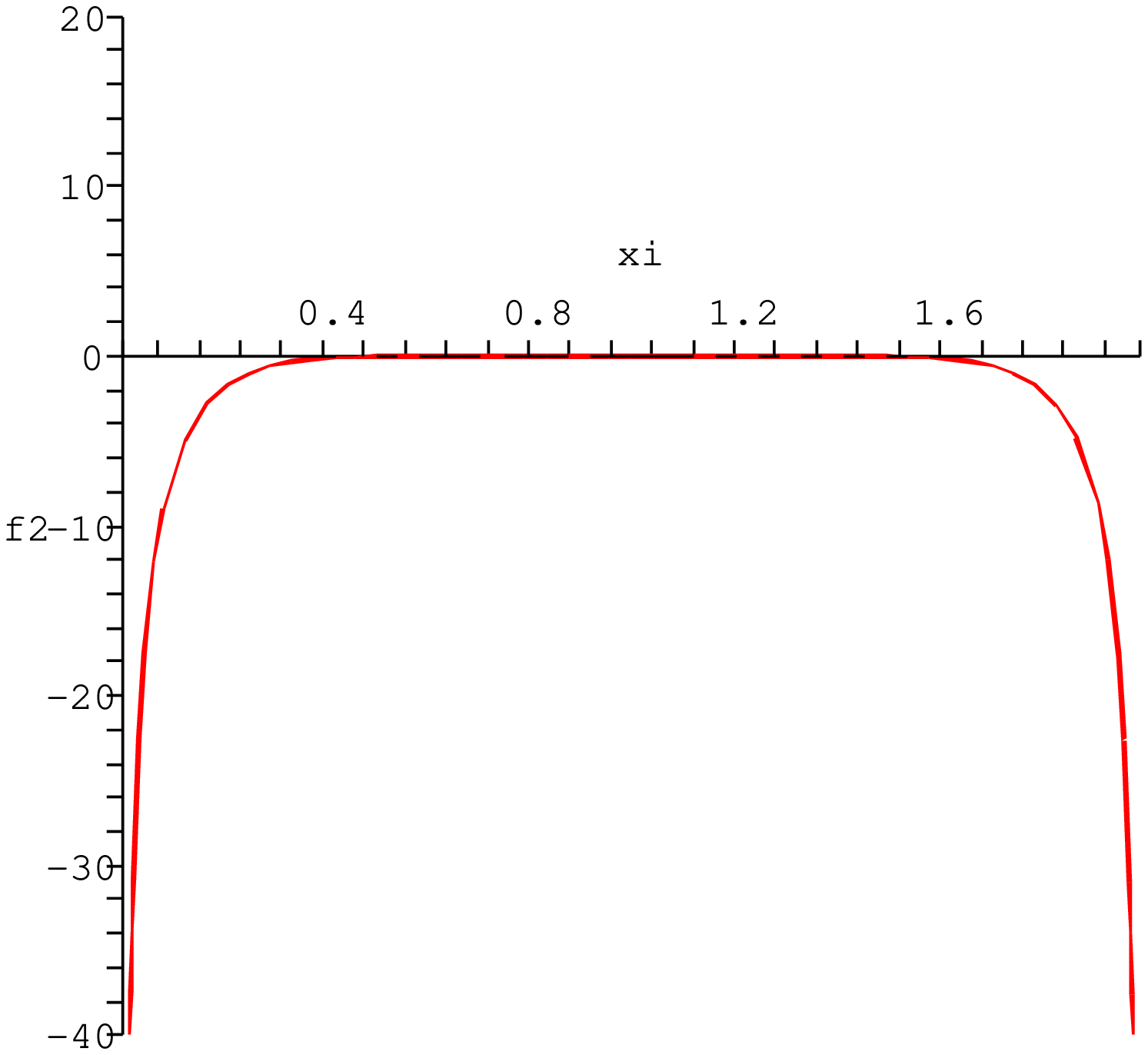,width=.35\textwidth}}
      \subfigure[]{\epsfig{figure=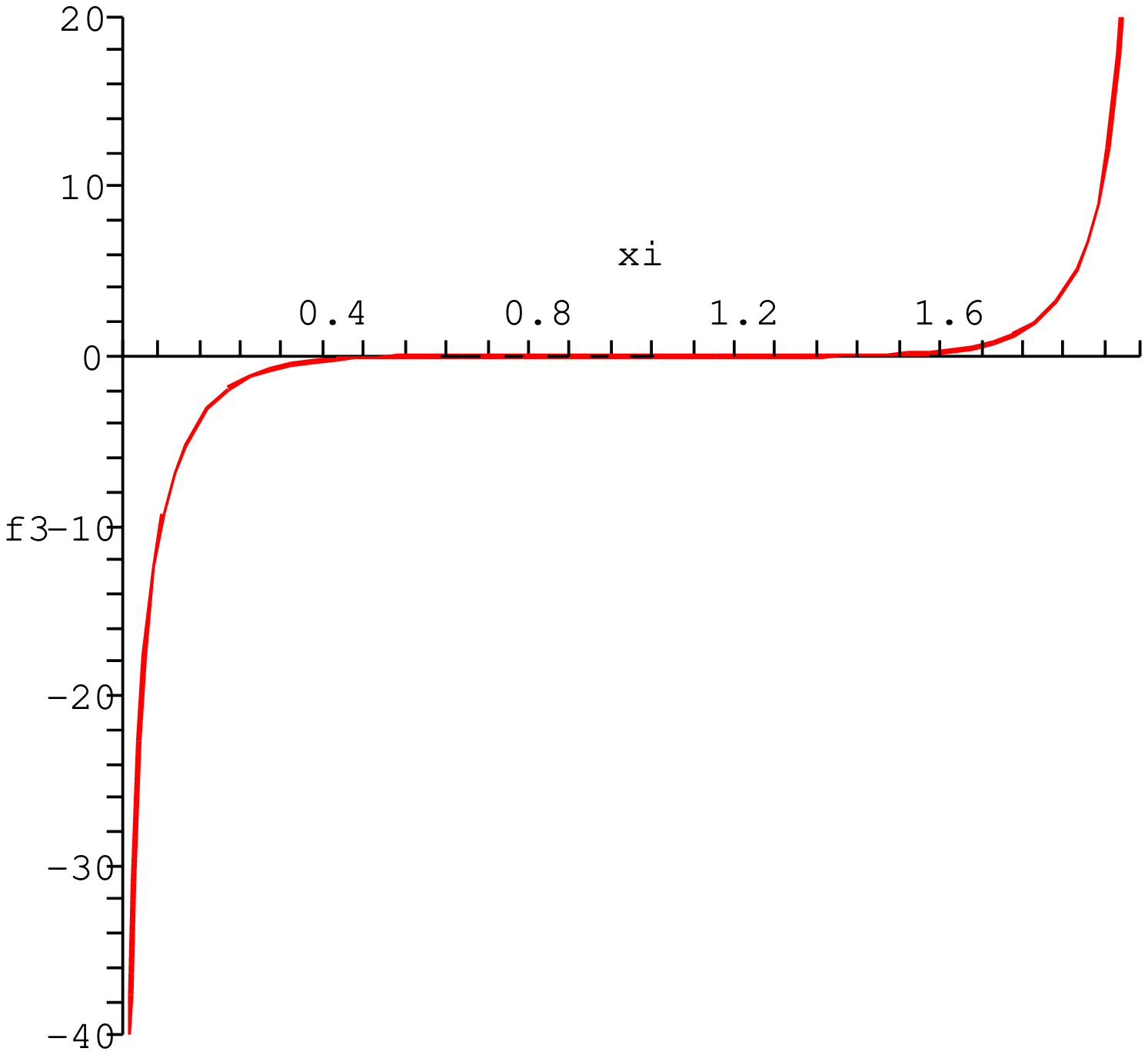,width=.35\textwidth}}}
\caption{Plots of the solutions (a) $f_1(\xi,k)$ (b) $f_2(\xi,k)$, (c) 
$f_3(\xi,k)$ with $k = 0.9999999999$}
\label{fig:fsolns}
\end{figure}

The appropriate solutions to Nahm's equations are \cite{Brown:1982gz}
\begin{eqnarray}
\label{eq:f1soln}
f_{1}(\xi,k) & = & \frac{-K(k)}{\textrm{sn}(K(k)\xi,k)} \,\, ,\\
\label{eq:f2soln}
f_{2}(\xi,k) & = & \frac{-K(k)\textrm{dn}(K(k)\xi,k)}{\textrm{sn}(K(k)\xi,k)} \,\, , \\
\label{eq:f3soln}
f_{3}(\xi,k) & = & \frac{-K(k)\textrm{cn}(K(k)\xi,k)}{\textrm{sn}(K(k)\xi,k)} \,\, .
\end{eqnarray}
where $K(k)$ is the complete elliptic integral of the first kind and
$\text{sn}(\xi,k)$, $\text{cn}(\xi,k)$ and $\text{dn}(\xi,k)$ are the Jacobian 
elliptic functions. See ref. \cite{Erdelyi} for a review of the properties of elliptic functions. Note that the solutions \eqref{eq:f1soln} - \eqref{eq:f3soln} have divergences at $\xi=0$ and $\xi=2$, which correspond 
to the positions of the D3-branes along the D-strings. As noted in section \ref{sec:rescaling}, we have therefore fixed the length of the D-strings to be $L=2$. The parameter $k$ is a modulus which runs from $k=0$ to $k=1$. The limit $k \to 1$ corresponds to the asymptotic limit in which the D-strings are far apart from one another.  Figure \ref{fig:fsolns} shows graphs of $f_1(\xi,k)$, $f_2(\xi,k)$ and $f_3(\xi,k)$ for $k = 0.999999999$. In this limit $f_1(\xi,k)\sim K(k)$, and $f_2(\xi,k)\sim f_3(\xi,k) \sim0$ (except of course near the ends of the string, $\xi=0$ and $\xi=2$, where all three functions have poles). At $k=0$ we have $f_1(\xi,k) = f_2(\xi,k)$, and this configuration corresponds to the monopole configuration which is symmetric in the $x^1$--$\,x^2$ plane, where the two monopoles coincide.

If we take the limit $\alpha' \to 0$ in \eqref{eq:BIaction4}, keeping $v$ fixed, then 
\begin{displaymath}
\alpha'^{2} \tilde{H} \to \frac{4}{v^{4}} \partial_{\xi}
(f_{1}f_{2}f_{3}) \,\, ,
\end{displaymath} 
and we can expand the square root in \eqref{eq:BIaction4} to get
\begin{align}
S = -T_{1} \frac{\alpha' v}{2} \int_{-\infty}^{\infty} \!\! dt \int_{0}^{2} \! d\xi  
& \, \left\{ \tilde{H} + \frac{1}{\partial_{\xi}(f_{1}f_{2}f_{3})}  \right. \left[ 
-\frac{1}{2} \left( (f'_{1} - f_{2}f_{3})^{2} + \cdots \right) \right.
\nonumber \\ 
& + \frac{1}{2v^{2}} \left( \partial_{t} (f_{1}f_{2}f_{3}) \right) ^{2} 
 \left. \left. + \frac{1}{2v^{2}} \left( (\dot{f}_{1} f'_{2} - \dot{f}_{2} 
f'_{1} ) ^{2} + \cdots \right) \right] \right\} \,\, .
\label{eq:BIaction5}
\end{align}

We can calculate the metric on moduli space for the solutions \eqref{eq:f1soln} - \eqref{eq:f3soln} using Manton's technique for slowly moving monopoles, described in ref. \cite{Manton:1982mp}. The motion is described by a geodesic in the moduli space of parameters. So we must allow the modulus $k$ to vary slowly with time, such that the Bogomol'nyi equations \eqref{eq:fprime} continue to hold. Then the action reduces to 
\begin{displaymath}
S = -T_{1} \frac{\alpha' v}{2} \int_{-\infty}^{\infty} \!\! dt \int_{0}^{2} \! d\xi \,
\left\{ \tilde{H}
+ \frac{1}{2v^{2}} (\dot{f}_{1}^{2} + \dot{f}_{2}^{2}+\dot{f}_{3}^{2})
\right\} \,\, ,
\end{displaymath}
which gives the correct metric on moduli space, the Atiyah-Hitchin metric restricted to the geodesic we are considering, since $\tilde{H}$ is a total 
derivative.

\subsection{Validity of the Born-Infeld Action}

In what we have done we have been using the Born-Infeld action. However, there are limitations to the Born-Infeld action - it is not very accurate when the geometry is highly curved, because higher derivatives of the fields have been ignored. So we expect the action \eqref{eq:BIaction5}, which we obtained from the Born-Infeld action, to become inaccurate near the ends of the D-strings, where the geometry does become curved. This suggests that the limit we proposed in section \ref{sec:low_energy}, in which we take the string length to be small, may make the situation worse, since we are taking to be small the region in which the Born-Infeld action is accurate.

In the light of this discussion our calculation does not seem very promising. However, we may be redeemed by the fact that, in our approximation, we are working close to the BPS solutions \eqref{eq:f1soln} - \eqref{eq:f3soln}. More precisely, we have taken
\begin{equation}
\label{eq:fdot_small}
\dot{f}_i \sim \varepsilon \,\, ,
\end{equation}
and
\begin{equation}
\label{eq:fprime_small}
(f'_1 - f_2 f_3) \sim \varepsilon \,\, , \quad (f'_2 - f_3 f_1) \sim \varepsilon \,\, , \quad (f'_3 - f_1 f_2) \sim \varepsilon \,\, ,
\end{equation}
where $\varepsilon$ is small. Since we expect the BPS solution to be a correct solution to the full equations of motion, it is reasonable to assume that motion in which the solutions remain close to the BPS solution is accurately described by the Born-Infeld action \eqref{eq:BIaction5}.

\section{The Action for D-Strings in a D3-Brane Background}
\label{sec:BI_D3brane}

In the previous section we were working with D-strings in a flat background. In this section we will consider the same scenario, but this time in a D3-brane background in order to take into account the effects of the back-reaction of the D-branes on the geometry. We will find that the Bogomol'nyi equations for static solutions are unchanged when we include a D3-brane background geometry, as was suggested in refs. \cite{Gauntlett:1999xz}, \cite{Ghoroku:1999bc} and \cite{Constable:1999ac}. We will also conjecture that, in the D3-brane background, the action describing motion `close' to the Bogomol'nyi solutions is identical to the equivalent action in flat background.

In order for the supergravity D3-brane background metric to be accurate we will need to take a large number of D3-branes. So we will consider the case with  D-strings stretched between two parallel stacks of D3-branes, each of which contains $N_3$ D3-branes. In this configuration we have a gauge group $SU(2N_3)$ on the D3-branes broken down to $SU(N_3) \times SU(N_3) \times U(1)$.

\subsection{The Born-Infeld Action}

For now we concentrate on the Born-Infeld action. We will investigate the contribution from the Chern-Simons action in section \ref{sec:chern-simons} below. We again use the non-abelian Born-Infeld action for D-strings, as given in \eqref{eq:BIaction1} in the previous section, which we restate here
\begin{equation}
\label{eq:D3-brane_metric}
S_{\rm{BI}} = -T_{1} \iint \! d\tau d\sigma \,\, \textrm{STr} \left( \frac
{e^{-\phi} \left( -D \right)^{1/2}}{\left(- \det
\left(E^{ij}\right)\right)^{1/2}} \right) \,\, .
\end{equation}
As before we take $F_{ab}=0$ and we set $\Phi^{4}=\cdots=\Phi^{8}=0$. However, this time
we wish to include the effects of the D3-branes on the geometry, so the
appropriate background metric is
\[
ds^{2}=-\frac{dt^{2}}{\sqrt{\mathcal{H}}}+\sqrt{\mathcal{H}}d\sigma^{2} +
\frac{(dx^{1})^{2}+(dx^{2})^{2}+(dx^{3})^{2}}{\sqrt{\mathcal{H}}} + 
\sqrt{\mathcal{H}}(dx^{m})^{2} \,\, ,
\label{eq:D3brane-metric}
\]
where $m=4, \ldots, 8$ and the harmonic function $\mathcal{H}$ is given by  
\[
\mathcal{H} = 1 + \frac{g_sN_{3}}{\pi}\frac{\alpha'^{2}}{\sigma^{4}} +
\frac{g_sN_{3}}{\pi}\frac{\alpha'^{2}}{\left| \sigma-L \right| ^{4}} \,\, .
\]
This metric corresponds to $N_3$ D3-branes positioned at $\sigma=0$, and $N_3$
positioned at $\sigma = L$. We also have $B_{\mu\nu}=0$, and for D3-branes the dilaton $\Phi$ is constant. 

Then
\[
E_{\mu \nu} = \rm{diag} \left( -\frac{1}{\sqrt{\mathcal{H}}}, \sqrt{\mathcal{H}},
\frac{1}{\sqrt{\mathcal{H}}}, \frac{1}{\sqrt{\mathcal{H}}}, 
\frac{1}{\sqrt{\mathcal{H}}} \right) \,\, ,
\]
giving
\[
\det \left( E^{ij} \right) = \mathcal{H}^{3/2} \,\, .
\]
Also
\[
D = \det \begin{pmatrix}
-\frac{1}{\sqrt{\mathcal{H}}} & 0 & \alpha' \partial_{0} \Phi^{1} & \alpha'
\partial_{0} \Phi^{2} & \alpha' \partial_{0} \Phi^{3} \\
0 & \sqrt{\mathcal{H}} & \alpha' \partial_{1} \Phi^{1} &  \alpha' \partial_{1}
\Phi^{2} &  \alpha' \partial_{1} \Phi^{3} \\
-\alpha' \partial_{0} \Phi^{1} & -\alpha' \partial_{1} \Phi^{1} &
\sqrt{\mathcal{H}} & \alpha' \left[ \Phi^{1}, \Phi^{2} \right] & \alpha' \left
[ \Phi^{1}, \Phi^{3} \right] \\
-\alpha' \partial_{0} \Phi^{2} & -\alpha' \partial_{1} \Phi^{2} &
\alpha' \left[ \Phi^{2}, \Phi^{1} \right] & \sqrt{\mathcal{H}} & \alpha' \left
[ \Phi^{2}, \Phi^{3} \right] \\
-\alpha' \partial_{0} \Phi^{3} & -\alpha' \partial_{1} \Phi^{3} &
\alpha' \left[ \Phi^{3}, \Phi^{1} \right] & \alpha' \left[ \Phi^{3},
\Phi^{2} \right] & \sqrt{\mathcal{H}} 
\end{pmatrix} \,\, .
\]

As in the previous section, to describe the D-strings funnelling out into 
D3-branes, we should take the fields $\Phi_{i}$ to be proportional to the Pauli
matrices
\begin{equation}
\label{eq:Phi_j}
\Phi_{j} = -\frac{1}{2} \phi_{j} \sigma_{j} \quad \textrm{(no summation over j)} \,\, ,
\end{equation}
where $j =1,2,3$, and the $\phi_{j}$ are real functions of $t$ and $\sigma$
Then some calculation yields
\begin{align}
S_{\rm{BI}} = -T_{1} \iint \! dt d\sigma \,\, & \left( \left( 1 +
\frac{\alpha'^{2}}{4\mathcal{H}} \partial_{\sigma} (\phi_{1}\phi_{2}\phi_{3}) 
\right)^{2} - \frac{\alpha'^{2}}{4}(\dot{\phi_{1}}^{2} +\dot{\phi_{2}}^{2} +
\dot{\phi_{3}}^{2}) \right.
\nonumber \\
& - \frac{\alpha'^{4}}{16\mathcal{H}} \left( \partial_{t}
(\phi_{1}\phi_{2}\phi_{3}) \right)^{2}  
 + \frac{\alpha'^{2}}{4\mathcal{H}} \left(
 (\phi_{1}'-\phi_{2}\phi_{3})^{2} + \cdots \right)
\nonumber \\
& \left. - \frac{\alpha'^{4}}{16\mathcal{H}} \left( (\dot{\phi_{1}} \phi_{2}' - \dot{\phi_{2}}
\phi_{1}')^{2} + \cdots \right) \right)^{1/2} \,\, .  
\end{align}

\subsection{Rescaling the Action}

To compare with the flat background case we will need to rescale the
coordinate $\sigma$ to a new coordinate $\xi$ so that one D3-brane is
positioned at $\xi=0$ and the other is at $\xi=2$. So we need to calculate
the coordinate distance between the two branes. In order to do this we consider the action for a test D-string in the supergravity background \eqref{eq:D3brane-metric}. The Born-Infeld action for the D-string is of the form
\[
S_{\rm{BI}} = - T_1 \iint \! d\sigma dt \, \sqrt{ -\det( \partial_{a} X^{\mu}
\partial_{b} X^{\nu} g_{\mu \nu})} \,\, ,
\]
where $g_{\mu \nu}$ is given by \eqref{eq:D3brane-metric}. We can take
\begin{eqnarray}
X^{0}  =  t \,\, ,  &
X^{9}  =  \sigma \,\, , \\
X^{i}  =  v^{i}t \,\, ,  &
X^{m}  =  \omega^{m}t \,\, ,
\end{eqnarray}
where $i=1,2,3$ and $m=4,\ldots,8$. Since the string must end on the
D3-brane, we have $\omega^{m}=0$.
So
\begin{align}
S_{\rm{BI}} & = -T_1 \iint \! d\sigma dt \, \sqrt{ -\det \begin{pmatrix}
    (1 - v^{i}v^{i}) \mathcal{H}^{-1/2} & 0 \\
    0 & \mathcal{H}^{1/2}
    \end{pmatrix} } \\
  & = -T_1 \iint \! d\sigma dt \, \sqrt{1-v^{i}v^{i}} \\
  & = -LT_1 \int\! dt \sqrt{1-v^{i}v^{i}} \,\, ,
\end{align}
where $L$ is the coordinate length along the string. Comparing this to
the action for a relativistic particle
\[
S = - m \int \! dt \, \sqrt{1 - (v^{i})^{2}} \,\, ,
\]
we obtain
\[
L = \frac{m}{T_1} = \alpha' v \,\, .
\]
This is identical to the coordinate length along the string in the
flat background case.

We rescale the coordinate $\sigma$ and the fields $\phi_{i}$ to $\xi$ and
$f_{i}$ respectively, so that the positions of the two D-branes
become 0 and 2, and the $f_{i}$ still obey Nahm's equations, as we did in section \ref{sec:rescaling} for the flat background case.
So
\begin{align}
\sigma & = \frac{\alpha' v }{2} \xi \,\, , \\
\phi_{i} & = \frac{2}{\alpha' v} f_{i} \,\, .
\end{align}
Then the action becomes
\begin{align}
S_{\rm{BI}} = -T_1 \frac{\alpha' v}{2} \int_{-\infty}^{\infty} \!\! dt \int_{0}^{2} \! d\xi & \, 
\left[ 
\left( 1 + \frac{1}{\mathcal{H}} \frac{4}{\alpha'^{2}v^{4}} \partial_{\xi}
  (f_{1}f_{2}f_{3}) \right)^{2}
- \frac{1}{v^{2}} ( \dot{f_{1}}^{2} + \dot{f_{2}}^{2} +
  \dot{f_{3}}^{2} ) \right. \nonumber \\
& \left.  - \frac{1}{\mathcal{H}} \frac{4}{\alpha'^{2}v^{6}} ( \partial_{t}
  (f_{1}f_{2}f_{3}) )^{2} 
 + \frac{1}{\mathcal{H}} \frac{4}{\alpha'^{2}v^{4}} \left( (f_{1}' - f_{2}
  f_{3})^{2} + \cdots \right) \right. \nonumber \\ \label{eq:D3_BIaction1}
& \left.  - \frac{1}{\mathcal{H}} \frac{4}{\alpha'^{2}v^{6}} \left( 
( \dot{f_{1}} f_{2}' - \dot{f_{2}} f_{1}')^{2} + \cdots \right)
\right]^{1/2} \,\, ,
\end{align}
with
\[
\mathcal{H} = 1 + \frac{16g_sN_{3}}{\pi} \frac{1}{\alpha'^{2}v^{4}\xi^{4}} + \frac{16g_sN_{3}}{\pi} \frac{1}{\alpha'^{2}v^{4}\left| \xi-2 \right| ^{4}} \,\, ,
\]
where $\xi = 2$ is the location of the second D3-brane.

According to refs. \cite{Gauntlett:1999xz} and \cite{Ghoroku:1999bc}, the Bogomol'nyi equations derived from \eqref{eq:D3_BIaction1} should be the same as those we calculated for the flat background case, which are given in equations \eqref{eq:fprime}. In order to verify this we will need to include the contribution from the Chern-Simons action, which we will discuss in the next section.

\subsection{The Chern-Simons Action}

\label{sec:chern-simons}

The non-abelian Chern-Simons action for the D-strings takes the form \cite{Myers:1999ps}
\begin{equation}
\label{eq:cs_action}
S_{\rm{CS}} = \mu_1 \iint \! d\tau \, d\sigma \, \textrm{STr} \left( P\,[e^{i\alpha'i_{\Phi}i_{\Phi}} \sum C^{(n)}]\,e^{\alpha'F} \right) \,\, ,
\end{equation}
where $P$ denotes the pullback to the D-strings' worldvolume, and $i_{\Phi}$ denotes the interior product taken with respect to the $\Phi^i$. Here $C^{(n)}$ are the Ramond-Ramond potentials, and $\mu_1$ is the Ramond-Ramond charge of the D-strings. Since D-branes are BPS objects we have the following condition relating the D-string charge and the D-string tension
\begin{equation}
\label{eq:BPS_condition}
T_1 = 2 \mu_1 \,\, ,
\end{equation}
where the factor of two is present in \eqref{eq:BPS_condition} because we have two D-strings. In the D3-brane background there is a non-zero $C^{(4)}$ field, which is given by
\begin{equation}
\label{eq:C4}
C^{(4)} = (\mathcal{H}^{-1}-1) \, dt \wedge dx^1 \wedge dx^2 \wedge dx^3 \,\, ,
\end{equation}
Substituting \eqref{eq:C4} into the action \eqref{eq:cs_action} results in a non-trivial interaction term, which is
\[
S_{\rm{CS}} = i \, \alpha' \mu_1 \iint \! d\tau \, d\sigma \, \textrm{STr} \left( P\,[ i_{\Phi}i_{\Phi} C^{(4)}] \right) \,\, .
\]
Now
\[
i_{\Phi}i_{\Phi} C^{(4)} = \frac{1}{2} \left[\Phi^j,\Phi^i\right] C^{(4)}_{ij\mu\nu} \, dx^{\mu} \wedge dx^{\nu} \,\, .
\]
So we find
\[
S_{\rm{CS}} = i \frac{\alpha'^2 }{2} \,\, \mu_1 \iint \! dt\, d\sigma \,\textrm{STr} \left( \partial_{\sigma}\Phi^i [\Phi^k,\Phi^j]\,C^{(4)}_{tijk} \right) \,\, .
\]
Substituting in the ansatz \eqref{eq:Phi_j} for the $\Phi^i$, and the expression \eqref{eq:C4} for $C^{(4)}_{tijk}$, and performing the rescalings \eqref{eq:rescale_sigma} and \eqref{eq:rescale_phi} we find
\[
S_{\rm{CS}} = \frac{4}{\alpha'v^3} \,\, \mu_1 \iint \! dt\, d\xi \left( \mathcal{H}^{-1} - 1\right) \partial_{\xi} \left( f_1f_2f_3 \right) \,\, .
\]

Using the expression for $\tilde{H}$ \eqref{eq:H}, and ignoring the total derivative term and the constant terms,  we find
\begin{equation}
\label{eq:cs_action_1}
S_{\rm{CS}} = \alpha' v \mu_1 \! \iint \! dt\, d\xi \left( \frac{\tilde{H}}{\mathcal{H}} \right) \,\, .
\end{equation}
And using the BPS condition \eqref{eq:BPS_condition}, this becomes
\begin{equation}
\label{eq:cs_action_2}
S_{\rm{CS}} = \frac{\alpha' v}{2} \, T_1 \iint \! dt\, d\xi \left( \frac{\tilde{H}}{\mathcal{H}} \right) \,\, .
\end{equation}

\subsection{The Bogomol'nyi Equations}

In the D3-brane background the total action is given by
\begin{equation}
\label{eq:D3_total_action}
S = S_{\rm{BI}} + S_{\rm{CS}} \,\, ,
\end{equation}
where $S_{\rm{BI}}$ and $S_{\rm{CS}}$ are given by \eqref{eq:D3_BIaction1} and \eqref{eq:cs_action_2} respectively. If we consider a static configuration with $\dot{f}_1 = \dot{f}_2 = \dot{f}_3$, which also satisfies the ansatz
\begin{equation}
\label{eq:D3_fprime}
f'_{1} = f_{2}f_{3} \,\, , \qquad
f'_{2} = f_{3}f_{1} \,\, , \qquad
f'_{3} = f_{1}f_{2} \,\, ,
\end{equation}
then we find that this configuration satisfies the equations of motion for \eqref{eq:D3_total_action}. We deduce that \eqref{eq:D3_fprime} are the Bogomol'nyi equations for the case with the D3-brane background. Note that they are identical to those we had in the flat background case \eqref{eq:fprime}, as we would expect from refs. \cite{Gauntlett:1999xz}, \cite{Ghoroku:1999bc} and \cite{Constable:1999ac}.

In fact, the Bogomol'nyi equations are given by \eqref{eq:D3_fprime}, irrespective of the form of $\mathcal{H}$. This means that the Bogomol'nyi equations are not affected by the way in which we minimally break the $SU(2N_3)$ symmetry; for example a string stretched between a single D3-brane on one side and a large stack of D3-branes on the other would have the same Bogomol'nyi equations. This suggests that the D-strings may only be aware of one D3-brane at each end - it makes no difference to them how many other D3-branes we stack together. In order to test this suggestion further we would need to analyse the boundary conditions of the Nahm data, using for example the techniques described in \cite{Chen:2002vb}.

\subsection{The Supergravity Limit}

In section \ref{sec:bogomolnyi}, in the flat background case, we expanded out the square root in the Born-Infeld action by taking the low energy limit $\alpha' \to 0$. However, if we compare the action we used for section \ref{sec:bogomolnyi}, which was \eqref{eq:BIaction3}, with the action \eqref{eq:D3_BIaction1} which we are currently dealing with, there is a significant difference. In \eqref{eq:BIaction3} the leading order term in $\alpha'$ is the first term under the square root, which is of order $1/\alpha'^4$. However, in \eqref{eq:D3_BIaction1} all terms are of leading order $\alpha'^0$, and so all contribute to the leading order term in $\alpha'$. This would make an expansion in $\alpha'$ very messy for the action \eqref{eq:D3_BIaction1}.

There is an alternative limit we can take; for the supergravity solution to be accurate we need $N_3$ to be large. In this limit
\begin{equation}
\label{eq:mathcal_H}
\frac{1}{\mathcal{H}} \sim \frac{1}{N_{3}}\alpha'^{2}v^{4}h(\xi) \,\, ,
\end{equation}
where $h(\xi)$ is a function of $\xi$ given by
\[
h(\xi) = \frac{\pi}{16 g_s} \frac{1}{\xi^{-4} + |\xi - 2|^{-4}} \,\, .
\] 

We can expand the square root in \eqref{eq:D3_BIaction1} as a series in $1/N_{3}$. So
defining
\[
J \equiv 1 - \frac{1}{v^{2}}\left(\dot{f_{1}}^{2} + \dot{f_{2}}^{2} +
\dot{f_{3}}^{2}\right) \,\, ,
\]
then
\begin{align}
S_{\rm{BI}} & = -T_1 \frac{\alpha' v}{2} \int_{-\infty}^{\infty} \! \! dt 
\int_{0}^{2} \! d\xi \, \sqrt{J} \, \left[ 1
+ \frac{8 h(\xi)}{JN_{3}}\,\partial_{\xi}(f_{1}f_{2}f_{3})
+ \frac{16h(\xi)^2}{JN_{3}^{2}} \left
  ( \partial_{\xi}(f_{1}f_{2}f_{3}) \right)^{2} \right. \nonumber \\
&  \qquad \qquad \left. - \frac{4h(\xi)}{JN_{3}v^{2}} \left( \partial_{t}
  (f_{1}f_{2}f_{3}) \right)^{2} 
+ \frac{4h(\xi)}{JN_{3}} \left( (f_{1}' - f_{2} f_{3})^{2} +
  \cdots \right) \right. \nonumber \\
&  \qquad \qquad \left. - \frac{4h(\xi)}{JN_{3}v^{2}} \left( 
(\dot{f_{1}} f_{2}' - \dot{f_{2}} f_{1}')^{2} + \cdots \right) \right]^{1/2} \nonumber \\
& \simeq -T_1 \frac{\alpha' v}{2} \int_{-\infty}^{\infty} \!\! dt 
\int_{0}^{2} \! d\xi \, \sqrt{J} \, \left( 1
+ \frac{4h(\xi)}{JN_{3}}\, \partial_{\xi} (f_{1}f_{2}f_{3})
- \frac{2h(\xi)}{JN_{3}v^{2}} \left( \partial_{t}
  (f_{1}f_{2}f_{3}) \right)^{2} \right. \nonumber \\
&  \qquad \qquad \left. + \frac{2h(\xi)}{JN_{3}} \left( 
(f_{1}' - f_{2}f_{3})^{2} + \cdots \right)
- \frac{2h(\xi)}{JN_{3}v^{2}} \left( (\dot{f_{1}}f_{2}' -
\dot{f_{2}}f_{1}') + \cdots \right) \right) \,\, ,
\label{eq:D3-BIaction1}  
\end{align} 
where we have used the expression for $\mathcal{H}$ \eqref{eq:mathcal_H} in \eqref{eq:D3-BIaction1} in order to make the dependence on $N_3$ explicit.

Since we will be dealing with solutions close to the Bogomol'nyi bound we will assume that
$\dot{f}_{i} \sim \varepsilon$, and all cyclic permutations of $(f_{1}' - f_{2}f_{3}) \sim \varepsilon$, where $\varepsilon$ is small (see \eqref{eq:fdot_small} and \eqref{eq:fprime_small}). So we can expand the factors of
$\sqrt{J}$ and $1/\sqrt{J}$ in powers of $\varepsilon$ to get
\begin{align}
S_{\rm{BI}} \simeq -T_1 \frac{\alpha' v}{2} \int_{-\infty}^{\infty} \!\!
 & dt \int_{0}^{2} \! d\xi \, \left[ 1 
+ \frac{1}{\mathcal{H}}\tilde{H}
- \frac{1}{2v^{2}} \left(1 -  \frac{1}{\mathcal{H}}\tilde{H} \right) (\dot{f}_{1}^{2} +
  \dot{f}_{2}^{2} + \dot{f}_{3}^{2})   \right. \nonumber\\
& \left. -  \frac{2}{\mathcal{H}\alpha'^{2}v^{6}} 
\left( \partial_{t} (f_{1}f_{2}f_{3}) \right)^{2}
+ \frac{2}{\mathcal{H}\alpha'^{2}v^{4}} \left( (f_{1}' - f_{2}f_{3}
  )^{2} + \cdots \right) \right. \nonumber \\
& \left. - \frac{2}{v^{2}} \frac{1}{\mathcal{H}\alpha'^{2}v^{4}} \left( (\dot{f}_{1}
  f_{2}' - \dot{f}_{2} f_{1}')^{2} + \cdots \right) \right] \label{eq:D3-BIaction2} \,\, ,
\end{align}        
where
\begin{equation}
\label{eq:H}
\tilde{H} \sim  \frac{4}{\alpha'^{2}v^{4}} \partial_{\xi} 
(f_{1}f_{2}f_{3}) \,\, ,
\end{equation}
as we had in \eqref{eq:H_tilde} in the flat background case. 

Adding in the Chern-Simons term \eqref{eq:cs_action_2}, we find that the total action is
\begin{align}
S_{\rm{BI}} \simeq -T_1 \frac{\alpha' v}{2} \int_{-\infty}^{\infty} \!\!
 & dt \int_{0}^{2} \! d\xi \, \left[ 1 
- \frac{1}{2v^{2}} \left(1 - \frac{1}{\mathcal{H}}\tilde{H} \right) (\dot{f}_{1}^{2} +
  \dot{f}_{2}^{2} + \dot{f}_{3}^{2})   \right. \nonumber\\
& \left. -  \frac{2}{\mathcal{H}\alpha'^{2}v^{6}} 
\left( \partial_{t} (f_{1}f_{2}f_{3}) \right)^{2}
+ \frac{2}{\mathcal{H}\alpha'^{2}v^{4}} \left( (f_{1}' - f_{2}f_{3}
  )^{2} + \cdots \right) \right. \nonumber \\
& \left. -  \frac{2}{\mathcal{H}\alpha'^{2}v^{6}} \left( (\dot{f}_{1}
  f_{2}' - \dot{f}_{2} f_{1}')^{2} + \cdots \right) \right] \,\, .\label{eq:D3-BIaction3}
\end{align} 

\subsection{Comparing with the Flat Background Case}

We will compare \eqref{eq:D3-BIaction3} to the result we had for the flat background case, which was
\begin{align}
S \simeq -T_1 \frac{\alpha' v}{2} \int_{-\infty}^{\infty} & \! dt 
\int_{0}^{2} \! d\xi \, \left[ \tilde{H}
+ \frac{2}{\tilde{H}\alpha'^{2}v^{4}} \left( (f_{1}' -
  f_{2}f_{3})^{2} + \cdots \right)
- \frac{2}{\tilde{H}\alpha'^{2}v^{6}} \left( \partial_{t}
  (f_{1}f_{2}f_{3}) \right)^{2} \right. \nonumber \\
& \left. - \frac{2}{\tilde{H}\alpha'^{2}v^{6}} \left( (\dot{f}_{1}
  f_{2}' - \dot{f}_{2} f_{1}')^{2} + \cdots \right) \right] \,\, .\label{eq:flat_BIaction1}
\end{align}
The two actions \eqref{eq:D3-BIaction3} and \eqref{eq:flat_BIaction1} agree up to a total derivative for solutions close to the Bogomol'nyi bound, providing that 
\begin{equation}
\label{eq:H_conjecture}
\mathcal{H} = \tilde{H} \,\, .
\end{equation}
 Now $\mathcal{H}$ is a harmonic function, and $\tilde{H}$ is expressed in
terms of elliptic functions, so clearly the above equality cannot hold
exactly. But $\mathcal{H}$ and $\tilde{H}$ are qualitatively similar in that they are both symmetric about $\xi=1$, and they have the same pole behaviour at $\xi=0$ and $\xi=2$. Recall that the Born-Infeld action is only an approximate action, so
it is possible that when the full action is used instead, the equality
above may be exact.

\section{Scattering D-Strings}

\label{sec:scattering}

Here we consider how to make the static solutions \eqref{eq:f1soln} - \eqref{eq:f3soln} time dependent in order to describe the scattering of the D-strings.

\subsection{Describing the Scattering}

We wish to initiate the motion of the D-strings in the limit where the 
D-strings are very far apart. Recall from section \ref{sec:bogomolnyi} that the required limit is $k\to 1$. In this limit we have $f_1\approx K(k)$, $f_2\approx 0$ and $f_3\approx 0$. Recall that $\Phi_1$, the field describing the positions of the D-strings in the $x^1$ direction, is given by $\Phi_1 = f_1\sigma_1$, and $\sigma_1$ has eigenvalues $\pm 1$. Therefore the D-strings' positions are at $\pm f_1$ in the $x^1$ direction, and 0 in the $x^2$ and $x^3$ directions. 

We describe the D-string scattering by decreasing $k$ to $k=0$ as time
increases. At $k=0$ we have  $f_1 = f_2$, and the D-strings can be thought of
as being at a minimum distance apart from each other (although we have to be
careful about what `distance' actually means in this scenario, since we are
dealing with noncommutative geometry). At this point we have rotational symmetry
in the $x^1$--$\,x^2$ plane, in correspondence with the symmetry found in monopoles
at minimum distance.

To conclude the scattering picture we swap the roles of $f_1$ and $f_2$ at $k=0$.
Therefore as time tends to infinity, the D2-strings grow further apart, but this
time in the $x^2$-direction. So the D-strings scatter at $90^{\circ}$, as we
would expect from the monopole point of view \cite{Atiyah:1988jp}.

\subsection{A Symmetry of the Solutions}

Under the transformation
\begin{displaymath}
k \mapsto \tilde{k} = \frac{ik}{k'} \,\, , \qquad
\xi \mapsto \tilde{\xi} = k' \xi  \,\, ,
\end{displaymath}
the Jacobian elliptic functions transform as 
\begin{displaymath}
\text{sn}(\tilde{\xi},\tilde{k})  =  k' \frac{\text{sn}(\xi,k)}
{\text{dn}(\xi,k)} \,\, , \quad
\text{cn}(\tilde{\xi},\tilde{k})  =  \frac{\text{cn}(\xi,k)}{\text{dn}(\xi,k)}
\,\, , \quad
\text{dn}(\tilde{\xi},\tilde{k})  =  \frac{1}
{\text{dn}(\xi,k)} \,\, .
\end{displaymath}
Also the complete elliptic integral of the first kind $K(k)$ transforms as
\begin{displaymath}
K(\tilde{k}) = k' K(k) \,\, ,
\end{displaymath}
(see for example ref. \cite{Erdelyi}). So the
functions $f_1$, $f_2$ and $f_3$ transform as
\begin{displaymath}
f_1(\tilde{\xi},\tilde{k})  =  f_2(\xi,k) \,\, , \quad
f_2(\tilde{\xi},\tilde{k})  =  f_1(\xi,k) \,\, , \quad
f_3(\tilde{\xi},\tilde{k})  =  f_3(\xi,k) \,\, .
\end{displaymath}
Therefore this transformation takes motion from before the scattering to motion
after the scattering and vice versa (since at the point of scattering $k=0$,
$f_1$ and $f_2$ are interchanged).

Under this transformation $k$ takes on imaginary values. Therefore we could
think of the motion after scattering as being described by a modulus
$k$ continued to imaginary values. Equivalently we could think of the motion as
being described by a modulus $k^2$, with $k^2 \to 1$ as the initial
configuration, and $k^2 \to -\infty$ as the final configuration, and the point
at which $k^2$ is zero as the `minimum distance' configuration.

\section{Perturbing the Fields}
\label{sec:perturbation}

We wish to describe the scattering of the two D-strings, taking into account
the possibility of energy radiation. Therefore it is not enough to allow the
modulus $k$ to depend on time in the solutions to the Bogomol'nyi
equations \eqref{eq:f1soln} - \eqref{eq:f3soln}; we also need to include
perturbations to account for the energy in the non-zero modes. We perturb the ansatz \eqref{eq:Phi} as follows
\begin{equation}
\label{eq:perturbed-ansatz}
\Phi^i = \sum_i \left(-\frac{1}{2}\phi_i(\sigma,k) + \tilde{\varepsilon}_i(\sigma,\tau)\right)\sigma_i + \sum_{i\neq j} \hat{\varepsilon}_{ij}(\sigma,\tau)\,\sigma_j \,\, ,
\end{equation}
where $\phi_i(\sigma,k)$ is the static solution to the Bogomol'nyi equations \eqref{eq:nahm} (it is related to the static solution \eqref{eq:f1soln} - \eqref{eq:f3soln} by a rescaling of the $\phi_i$ given by equation \eqref{eq:rescale_phi}, and a rescaling of $\sigma$ given by \eqref{eq:rescale_sigma}). The $\tilde{\varepsilon}_i(\sigma,\tau)$ can be thought of as `diagonal' perturbations of the $\Phi^i$, with the $\hat{\varepsilon}_{ij}(\sigma,\tau)$ as the `off-diagonal' perturbations. As before, the $\sigma_i$ are the Pauli matrices. Substituting the ansatz \eqref{eq:perturbed-ansatz} into the non-Abelian Born-Infeld action \eqref{eq:BIaction1}, and applying the symmetrised trace prescription, we find that there are no terms in the action which are of linear order in the $\hat{\varepsilon}_{ij}$. This means that the equations of motion for the $\hat{\varepsilon}_{ij}$ are at least linear in $\hat{\varepsilon}_{ij}$, and so the $\hat{\varepsilon}_{ij}$ can be consistently set to zero. Furthermore, when we evolve a configuration with $\hat{\varepsilon}_{ij}=0$ initially the $\hat{\varepsilon}_{ij}$ modes can never be excited. When we calculate the energy radiated during D-string scattering we will use an initial configuration tangent to the static solution, i.e. $\Phi^i(\xi,t=0) = -\phi_i(\sigma,k_0)\sigma_i/2$, where $k_0$ is some initial value of $k$, so we will have $\hat{\varepsilon}_{ij}=0$ initially. Therefore, since we will have $\hat{\varepsilon}_{ij}=0$ initially, we should take $\hat{\varepsilon}_{ij}=0$ at all times.

We have shown that we can neglect the `off-diagonal' perturbations of the $\Phi_i$ when we perturb the static solution. Therefore, instead of working from scratch with the non-Abelian Dirac-Born-Infeld action \eqref{eq:BIaction1}, we can just perturb the fields in the low energy, rescaled action \eqref{eq:BIaction4}. We will relabel the fields $f_i(\xi,t)$ in that action as $\varphi_i(\xi,t)$ in order to keep the notation $f_i$ for the static solutions \eqref{eq:f1soln} - \eqref{eq:f3soln}. Then we write
\begin{equation}
\label{eq:perturbedphi}
\varphi_{i} = f_{i} + \varepsilon_{i} \,\, ,
\end{equation}
We take the slow motion approximation,
with $\dot{f}_i$ small, and therefore we can assume that the perturbations
$\varepsilon_i$ and their derivatives $\varepsilon'_i$ and $\dot{\varepsilon}_i$
are also small and of the same order as $\varepsilon_i$. Our results will differ from those found in ref. \cite{Constable:1999ac} because in that paper they assumed a spherically symmetric ansatz for the D3-brane fields.

In order to find the equations of motion for the perturbations $\varepsilon_i$
we substitute \eqref{eq:perturbedphi} into the action \eqref{eq:BIaction1} to 
get
\begin{align}
S =  -T_1 \frac{\alpha' v}{2} \int_{-\infty}^{\infty} \!\!\! dt \, \int_{0}^{2} \! d\xi \,\,\, 
\frac{1}{2} \,\, & \bigg[ 
 \left\{ (\dot{f}_{1} + \dot{\varepsilon}_{1})^{2}
+ (\dot{f}_{2} + \dot{\varepsilon}_{2})^{2}
+ (\dot{f}_{3} + \dot{\varepsilon}_{3})^{2} \right\} \nonumber \\
&- \, \frac{1}{H} \,\, \Big\{ 
(\varepsilon'_{1} - f_{2} \varepsilon_{3} - f_{3} \varepsilon_{2})^{2} 
+ \cdots \Big\} \bigg] \,\, ,
\label{eq:BIaction6}
\end{align}
where $+ \cdots$ again denotes the addition of all cyclic permutations of the indices of the first term
in the brackets, and we have defined
\begin{displaymath}
H = (f_{1}^{2} f_{2}^{2} + f_{2}^{2} f_{3}^{2} + f_{3}^{2} f_{1}^{2}) \,\, .
\end{displaymath}

We next define the row vectors $O_i$,
\begin{eqnarray}
O_{1} & = & \left( \begin{array}{ccc} \partial_{\xi} & -f_{3} & -f_{2} 
\end{array} \right) \,\, , \nonumber \\ 
O_{2} & = & \left( \begin{array}{ccc} -f_{3} & \partial_{\xi} & -f_{1} 
\end{array} \right) \,\, ,\nonumber \\ 
O_{3} & = & \left( \begin{array}{ccc} -f_{2} & -f_{1} & \partial_{\xi} 
\end{array} \right) \,\, .\nonumber 
\end{eqnarray}
Then we can rewrite the action \eqref{eq:BIaction6} in the more compact form
\begin{equation}
\label{eq:BIaction7}
S = -T_1 \frac{\alpha' v}{2} \int_{-\infty}^{\infty} \!\! dt \int_{0}^{2} \! d\xi \, \bigg[
\frac{1}{2} (\vec{\dot{f}} + \vec{\dot{\varepsilon}})^{2} - \frac{1}{2H} (O_{i}
\vec{\varepsilon}, O_{i} \vec{\varepsilon}\,) \bigg] \,\, .
\end{equation}

If we had taken the alternative limit in section \ref{sec:low_energy}, i.e. $v \to \infty$ as $\alpha' \to 0$, which is the Yang-Mills limit, then we would have $H=1$ in \eqref{eq:BIaction7}. Then the equations of motion with respect to
$\vec{\varepsilon}$ would be 
\begin{displaymath}
\vec{\ddot{f}} + \vec{\ddot{\varepsilon}} + O_{i}^{\dag} O_{i} \vec{\varepsilon}
= 0 \,\, ,
\end{displaymath}
where
\begin{displaymath}
O_{1}^{\dag} = \left( \begin{array}{c} -\partial_{\xi} \\ -f_{3} \\ -f_{2}
\end{array} \right)
\,\, , \quad 
O_{2}^{\dag} = \left( \begin{array}{c} -f_{3} \\ -\partial_{\xi} \\ -f_{1}
\end{array} \right)
\,\, , \quad 
O_{3}^{\dag} = \left( \begin{array}{c} -f_{2} \\ -f_{1} \\ -\partial_{\xi}
\end{array} \right)
\,\, .
\end{displaymath}

Now we include the factor of $H$ in \eqref{eq:BIaction7}. 
Following the method from ref. \cite{Callan:1998kz}, we rescale $\xi \mapsto
x$ such that
\begin{equation}
\label{eq:dsigmadx}
\frac{dx}{d\xi} = \sqrt{H} \,\, ,
\end{equation}
and we make the following definitions
\begin{eqnarray}
\label{eq:Fi}
F_{i} & = & \frac{f_{i}}{\sqrt{H}} \,\, , \\
\label{eq:Omega1}
\Omega_{1} & = & \left( \begin{array}{ccc} \partial_{x} & -F_{3} & -F_{2}
\end{array} \right) \,\, , \\
\label{eq:Omega2}
\Omega_{2} & = & \left( \begin{array}{ccc} -F_{3} & \partial_{x} & -F_{1}
\end{array} \right) \,\, ,\\
\label{eq:Omega3}
\Omega_{3} & = & \left( \begin{array}{ccc} -F_{2} & -F_{1} & \partial_{x}
\end{array} \right)  \,\, .
\end{eqnarray}
Then substituting \eqref{eq:dsigmadx}, \eqref{eq:Fi}, \eqref{eq:Omega1},
\eqref{eq:Omega2} and \eqref{eq:Omega3} into \eqref{eq:BIaction7} we get
\begin{equation}
\label{eq:BIaction8}
S = -T_1 \frac{\alpha' v}{2}  \int_{-\infty}^{\infty} \! dt \int_{-\infty}^{\infty}
\frac{dx}{\sqrt{H}} \, \bigg[
\frac{1}{2} (\sqrt{H} \vec{\dot{F}} + \vec{\dot{\varepsilon}})^{2} 
- \frac{1}{2} (\Omega_{i} \vec{\varepsilon}, \Omega_{i} \vec{\varepsilon}\,) 
\bigg] \,\, .
\end{equation}
Next we rescale
\begin{equation}
\label{eq:varepsilon}
\varepsilon_{i} \mapsto \eta_{i} = \varepsilon_{i}\, H^{-1/4} \,\, ,
\end{equation}
and we redefine the $\Omega_{i}$
\begin{eqnarray}
\label{eq:Omega4}
\Omega_{1} & = & \left( \begin{array}{ccc} \partial_{x}(\ln{H^{1/4}}) 
+ \partial_{x} & -F_{3} & -F_{2}
\end{array} \right) \,\, , \\
\label{eq:Omega5}
\Omega_{2} & = & \left( \begin{array}{ccc} -F_{3} & \partial_{x}
(\ln{H^{1/4}}) + \partial_{x} & -F_{1}
\end{array} \right) \,\, , \\
\label{eq:Omega6}
\Omega_{3} & = & \left( \begin{array}{ccc} -F_{2} & -F_{1} & \partial_{x}
(\ln{H^{1/4}}) + \partial_{x}
\end{array} \right)  \,\, .
\end{eqnarray}
Then substituting \eqref{eq:varepsilon}, \eqref{eq:Omega4}, \eqref{eq:Omega5}
and \eqref{eq:Omega6} into \eqref{eq:BIaction8} we get
\begin{equation}
\label{BIaction9}
S = -T_1 \frac{\alpha' v}{2} \int_{-\infty}^{\infty} \! dt \int_{-\infty}^{\infty} \! dx \, \bigg[
\frac{1}{2} ( H^{1/4} \vec{\dot{F}} + \vec{\dot{\eta}})^{2} 
- \frac{1}{2} (\Omega_{i} \vec{\eta}, \Omega_{i} \vec{\eta}) 
\bigg] \,\, .
\end{equation}

Then the equation of motion for $\vec{\eta}$ is
\begin{equation}
\label{eq:etaeqn}
\vec{\ddot{\eta}} + \Omega_{i}^{\dag} \Omega_{i} \vec{\eta}
= -H^{1/4} \vec{\ddot{F}} - \frac{1}{4H^{3/4}}\dot{H}\vec{\dot{F}} \,\, .
\end{equation}
These equations are three coupled equations, with each of the form of a Laplacian with a potential given by $\Omega_i^\dag \Omega_i$, and a forcing term given by the right-hand side of equation \eqref{eq:etaeqn}.

\section{Conclusions}
\label{sec:conclusions}

We have presented the solutions to Nahm's equations \eqref{eq:f1soln} - \eqref{eq:f3soln} which represent D-strings stretched between D3-branes. We have descibed the slow motion scattering of the D-strings by allowing the modulus $k$ to depend on time. We have also derived the equations of motion \eqref{eq:etaeqn} for the time evolution of perturbations on the D-strings. 

We have already stated that our object in these calculations has been to calculate the energy radiated during the scattering of D-strings stretched between D3-branes. The perturbations corresponding to the non-zero modes of the solutions describe the energy radiated. Therefore it remains to solve the equations of motion for the perturbations and to calculate the energy retained in them in order to find out the energy radiated. We have performed this calculation numerically, and we will present the results in a following paper (ref. \cite{Barrett}).

A possible direction for future research would be to analyse the boundary conditions in the Nahm data corresponding to different ways of breaking the $SU(2N_3)$ symmetry. This would indicate whether or not the D-strings are indeed unaware of the D3-branes in the parallel stacks to which they are not attached.

\section*{Acknowledgements}

This work was partly supported by the EC network ``EUCLID", contract number HPRN-CT-2002-00325. JKB is supported by an EPSRC studentship. We thank Douglas Smith and Clifford Johnson for useful discussions.

\bibliographystyle{utcaps}
\bibliography{D3-D1Brane_paper_2_new}

\end{document}